# Channelized analog microwave short-time Fourier transform in the optical domain with improved measurement performance

Xiaowei Li,[1,2,†] Taixia Shi,[1,2,†] Dong Ma,[1,2] and Yang Chen[1,2,*]

[1] Shanghai Key Laboratory of Multidimensional Information Processing, East China Normal University, Shanghai, 200241, China
[2] Engineering Center of SHMEC for Space Information and GNSS, East China Normal University, Shanghai, 200241, China
*ychen@ce.ecnu.edu.cn
† These authors contributed equally to this paper

**Abstract:** In this article, analog microwave short-time Fourier transform (STFT) with improved measurement performance is implemented in the optical domain by employing stimulated Brillouin scattering (SBS) and channelization. By jointly using three optical frequency combs and filter- and SBS-based frequency-to-time mapping (FTTM), the time-frequency information of the signal under test (SUT) in different frequency intervals is measured in different channels. Then, by using the channel label introduced through subcarriers after photodetection, the obtained low-speed electrical pulses in different channels mixed in the time domain are distinguished and the time-frequency information of the SUT in different channels is respectively obtained and spliced to implement the STFT. For the first time, channelization measurement technology is introduced in the STFT system based on frequency sweeping and FTTM, greatly reducing the frequency-sweep range of the required frequency-sweep signal to the analysis bandwidth divided by the number of channels. In addition, channelization can also be used to improve the time and frequency resolution of the STFT system. A proof-of-concept experiment is performed. 12-GHz and 10-GHz analysis bandwidth is implemented by using a 4-GHz frequency-sweep signal and 3 channels and a 2-GHz frequency-sweep signal and 5 channels. Measurement performance improvement is also demonstrated.

## 1. Introduction

Accurate sensing of microwave frequencies is an important function of electronic warfare systems [1-3]. At present, the commonly used electromagnetic spectrum sensing methods in the electrical domain are mainly based on scanning superheterodyne [4]. By sequentially scanning the whole spectrum range of interest and down-converting it to the intermediate-frequency (IF) band, the frequency information of the signal under test (SUT) within the entire bandwidth can be sequentially obtained through digital signal processing in the IF band. This method has the characteristics of high accuracy, but the real-time performance of the measurement is severely limited by sweeping speed and accumulation time in the digital domain, making it difficult to adapt to application scenarios with high real-time requirements.

Besides the spectrum sensing that obtains the one-dimensional frequency information of the SUT, the two-dimensional time and frequency acquisition can also be used for spectrum sensing, which can provide not only the frequency information but also the relationship between frequency information and time. In the electrical domain, the two-dimensional time and frequency information is commonly obtained digitally using digital signal processing, such as short-time Fourier transform (STFT) [5, 6] and wavelet transform [7, 8]. However, because of its implementation in the digital domain, these methods have obvious problems when facing large bandwidth microwave applications: 1) It is difficult to sample high-frequency and large bandwidth microwave signals; 2) Ultra-high-speed data streams after sampling are difficult to process in real-time, and the time-frequency analysis has poor real-time performance.

Therefore, electrical domain spectrum sensing methods, whether based on scanning superheterodyne frequency measurement or time-frequency analysis using digital signal processing, is difficult to achieve high real-time measurement of large bandwidth signals. However, the high accuracy of frequency measurement is still extremely attractive. To overcome the bottleneck of traditional electrical domain spectrum sensing technology in real-time performance, there is an urgent need for new technology assistance: that is, using this new technology to achieve "rough" positioning of signal time and frequency within a wide bandwidth range, and then using traditional electronic technology to further accurately measure the signal within the "rough" frequency range. Because the frequency range to be scanned and measured by the electrical measurement method is greatly reduced by the "rough" frequency positioning, the real-time performance of the whole measurement will be greatly improved.

One of the promising new technologies that can assist traditional electrical domain spectrum sensing is microwave photonic signal measurement. Taking the distinguished advantages offered by microwave photonics, photonics-assisted one-dimensional microwave frequency sensing has been widely studied during the past few years [9–17]. Among these methods, filter- and SBS-based frequency-to-time mapping (FTTM) has received widespread attention due to its comprehensive performance in frequency measurement bandwidth, multi-frequency measurement capability, and measurement accuracy [14–17]. In addition to the one-dimensional frequency measurement, photonics-assisted two-dimensional time and frequency acquisition methods have also been investigated recently for obtaining more comprehensive information about SUT. STFT is one of the time and frequency acquisition methods, and several STFT methods realized in the optical domain have been reported [18–20]. However, all the methods are implemented by using large dispersion mediums, which are indeed achieved by adding a time dimension to dispersion-based FTTM in one-dimensional frequency measurement [9, 10]. The dispersion medium in these systems limits the analysis bandwidth and reconfigurability of these STFT methods. To break the limitation of the dispersion medium in the photonics-assisted time and frequency acquisition methods, we have conducted a series of research work based on the filter- and SBS-based FTTM by adding a time dimension through periodic high-speed frequency sweeping, and reported the first STFT system [21] and wavelet-like transform system [22] using the filter- and SBS-based FTTM. To avoid the use of high-speed frequency-sweep electrical sources used in [21, 22], a method without using high-frequency electrical devices and equipment was reported [23]. However, in [21–23], the analysis bandwidth of the system is still limited by the sweeping bandwidth of the electrical or optical frequency-sweep signal. In addition, as discussed in [21], the bandwidth of the frequency-sweep signal ultimately limits the time and frequency resolution of the system. Although we have demonstrated a novel method to improve the frequency resolution by broadening the SBS gain bandwidth [24], the simultaneous improvement of system analysis bandwidth, frequency resolution, and time resolution is still an urgent problem to be solved.

In this article, analog microwave STFT with improved measurement performance is implemented in the optical domain. By jointly using three optical frequency combs (OFCs) and filter- and SBS-based FTTM, the time-frequency information of the SUT in different frequency intervals is measured in different channels. Then, by using the channel label introduced through subcarriers after photodetection, the obtained low-speed electrical pulses in different channels mixed in the time domain are distinguished and the time-frequency information of the SUT in different channels is respectively obtained and spliced to implement the STFT. To the best of our knowledge, this is the first time that channelization measurement technology is introduced in the STFT system based on frequency sweeping and FTTM, which greatly reduces the frequency-sweep range of the required frequency-sweep signal to the analysis bandwidth divided by the number of channels. In addition, channelization can also be used to improve the time and frequency resolution of the STFT system. A proof-of-concept experiment is performed. 12-GHz and 10-GHz analysis bandwidth is implemented by using a 4-GHz frequency-sweep signal and 3 channels and a 2-GHz frequency-sweep signal and 5 channels. Measurement performance improvement is also demonstrated.

## 2. Principle and system

### 2.1 Channelized analog microwave STFT system

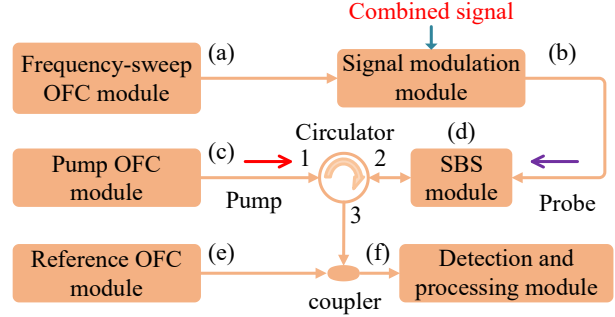

Fig. 1. Schematic diagram of the proposed channelized analog microwave STFT system.

Fig. 1 is the schematic diagram of the proposed channelized analog microwave STFT system in the optical domain. The key feature of the system is the employment of three different OFCs. The linear frequency-sweep OFC is generated by the frequency-sweep OFC module, which is then carrier-suppressed right single-sideband (CS-RSSB) modulated by a combined signal at the signal modulation module. The combined signal is comprised of the SUT and a fixed-frequency reference signal. The signal after the CS-RSSB modulation is then launched into an SBS module. The pump OFC generated by the pump OFC module is reversely injected into the SBS module via an optical circulator to simultaneously excite multiple stimulated Brillouin gains, which amplify the signal after the CS-RSSB modulation at certain wavelengths. The optical signal after SBS amplification is then combined with a reference OFC generated from a reference OFC module. Then, the combined optical signal from the optical coupler is detected and converted to electrical pulses carried by different subcarriers. By using the channel label introduced through subcarriers after photodetection, the obtained low-speed electrical pulses in different channels mixed in the time domain are distinguished and the time-frequency information of the SUT in different channels are respectively obtained and spliced to implement the STFT. Note that the reference signal is used to determine the relative position and corresponding frequency information of the pulses generated by the SUT

The three OFCs in Fig. 1 can be implemented in multiple ways. For example, Kerr combs were generated by pumping a high-Q microresonator using a continuous wave (CW) light wave [25]. In [26], by generating optical oscillation in a high-index silicon-glass microring resonator, an optical frequency comb was generated. In these works, Kerr combs with more than 70 spectral lines can be generated, and the spectral line spacing ranges from 200 GHz to more than 6 THz. Currently, the Kerr combs cannot be produced in our laboratory, so we choose another method to generate the OFCs used in Fig. 1, which is implemented by external modulation of optical modulators [27–30]. The basic principle of this kind of method is to generate equidistant RF-modulated optical sidebands by applying an external RF signal, thereby achieving the generation of OFCs. The main advantage of this kind of OFC is that it can achieve flexible tuning of the spectral line spacing by changing the frequency of the RF signal. In this article, proof-of-concept experiments are conducted using OFCs with 3 or 5 comb lines, so a CW laser diode (LD) and a single modulator injected by an RF signal are used as the pump/reference

OFC module. In the frequency-sweep OFC module, the OFC generated by the above structure is further carrier-suppressed left single-sideband (CS-LSSB) modulated by a frequency-sweep signal in a dual-parallel Mach–Zehnder modulator (DP-MZM) to generate the frequency-sweep OFC.

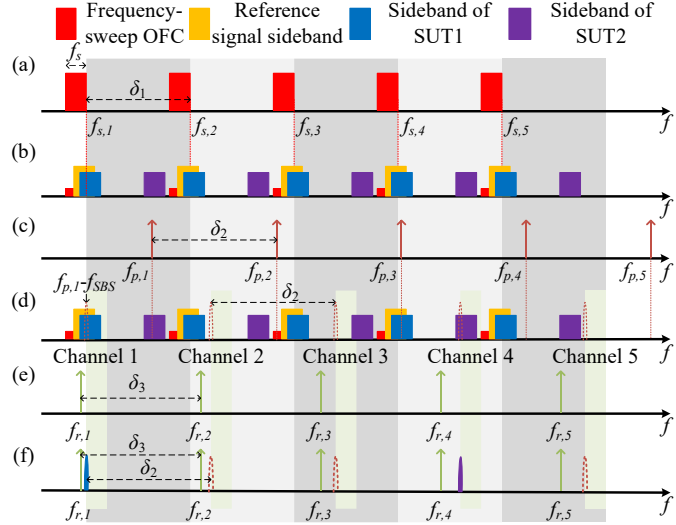

Fig. 2. Spectral diagrams at different positions in Fig. 1.

Taking the spectral line number of the OFC $N = 5$ as an example, the principle of implementing STFT in the system is shown in Fig. 2. Fig. 2(a) shows the spectrum of the linear frequency-sweep OFC, which has a sweep bandwidth of $f_s$, a sweep chirp rate of $k$, a sweep period of $T$, comb frequencies of $f_{s,n}$ ($n = 1, 2, 3, …, N$) and a free spectral range (FSR) of $\delta_1$. Note that $f_s$ equals to the instantaneous analysis bandwidth of each channel, $N$ also represents the number of channels in the system, and $\delta_1$ is indeed the instantaneous analysis bandwidth $B$ of the system. The three parameters have the following relationship

$$\delta_1 = B = Nf_s. \tag{1}$$

The linear frequency-sweep OFC shown in Fig. 2(a) is input to the signal modulation module in Fig. 1. The signal modulation module loads the SUT and the fixed-frequency reference signal $f_r$ onto the linear frequency-sweep OFC through optical CS-RSSB modulation. Here, two single-tone signals with different frequencies are used as the SUTs for the convenience of analysis. Due to multiple linear frequency-sweep comb lines, after CS-RSSB modulation, the reference signal and the SUTs generate optical sidebands on the right side of each frequency-sweep comb line, achieving the replication of the SUTs. Thus, $N$ channels are obtained. The optical spectrum after CS-RSSB modulation is shown in Fig. 2(b). Because of the CS-RSSB modulation, the red carrier is greatly suppressed.

The spectrum of the pump OFC is shown in Fig. 2(c). The first spectral line of the pump OFC is separated from the linear frequency-sweep OFC by the Brillouin frequency shift. The comb frequency relationship between the frequency-sweep OFC and the pump OFC is expressed as

$$f_{s,n} = f_{p,n} - (n-1)f_s - f_{SBS}, \tag{2}$$

where $f_{SBS}$ is the Brillouin frequency shift, $f_{p,n}$ ($n$ = 1, 2, 3, …, $N$) is the comb frequency of the pump OFC, $f_{s,n}$ ($n$ = 1, 2, 3, …, $N$) is the comb frequency of the frequency-sweep OFC, $\delta_2$ is the FSR of the pump OFC. $\delta_2$ is greater than $\delta_1$ by $f_s$ and can be expressed as

$$\delta_2 = \delta_1 + f_s. \qquad (3)$$

As shown in Fig. 2(d), due to the precisely designed OFC comb frequencies of the two OFCs, the multiple Brillouin gains excited by the pump OFC are located at different positions in $N$ different channels, and can respectively cover the optical sideband ranges generated by SUTs in $N$ different frequency bands (0 ~ $f_s$, $f_s$ ~ 2 $f_s$, 2 $f_s$ ~ 3 $f_s$ … ($N-1$) $f_s$ ~ $Nf_s$), that is

$$f_{s,n} + (n-1)f_s \sim f_{s,n} + nf_s, \qquad (4)$$

where $n$ = 1, 2, 3, …, $N$.

In Fig. 2, two single-tone signals are used as the SUTs. It can be seen that the frequency-sweep optical sidebands of the reference signal and SUT1 will pass through the Brillouin gain in the first channel so that part of the sidebands will be amplified. In other replications of these two sidebands, there is no interaction with other Brillouin gains of the rest channels. SUT2 has a relatively higher frequency, so its sideband is farther away from the sweep carrier than that of the reference signal and SUT1. As can be seen, only the fourth replication of it is amplified by the Brillouin gain in the fourth channel. If each channel of the optical signal selectively amplified by the SBS gains as shown in Fig. 2(d) is selected through optical filtering and then separately detected in a photodetector (PD), the time-frequency information within the corresponding frequency range of each channel can be obtained separately by processing the low-speed electrical pulses generated in the time domain [21]. However, accurate and ideal optical filtering is difficult to achieve, and parallel detection and processing can bring high complexity.

If the optical signals after selected amplification by the SBS gain in Fig. 2(d) are simply detected in a PD without optical processing, the electrical pulses in the time domain from different channels cannot be distinguished. To solve this problem, a reference OFC as shown in Fig. 2(e) is introduced, which is combined with the above-discussed optical signal after SBS interaction. $f_{r,n}$ ($n$ = 1, 2, 3, …, $N$) is the comb frequency of the reference OFC, $\delta_3$ is the FSR of the reference OFC. In this case, as shown in Fig. 1, the combined optical signal in Fig. 2(f) is directly detected. Because of the introduction of the reference OFC, each selectively amplified optical spectrum in Fig. 2(f) has a pure optical local oscillator close to it. Therefore, after photodetection, the generated electrical pulses in different channels are carried by different subcarriers. By properly designing the reference OFC and using the subcarriers as channel labels, the obtained low-speed electrical pulses in different channels mixed in the time domain can be distinguished. Thus, the time-frequency information of the SUT in different channels can be respectively obtained and spliced to implement the STFT.

The relationship between $\delta_3$ and $\delta_2$ is designed to be

$$\delta_3 = \delta_2 - \Delta f, \qquad (5)$$

where $\Delta f$ is the frequency difference of subcarriers in the adjacent channels. The relationship between $f_{p,n}$ and $f_{r,n}$ can be expressed as

$$f_{r,n} = f_{p,n} - f_{SBS} - f_{sc,min} - (n-1)\Delta f, \qquad (6)$$

where $f_{sc,min}$ is the minimum frequency of the subcarriers.

In the example shown in Fig. 2, due to the selection of two single-tone signals as the SUT, filtered optical pulses, as well as the finally generated low-speed electrical pulses carried by subcarriers, are only generated in channels 1 and 4. The subcarrier frequencies in different channels can be written as

$$f_{sc,n} = f_{sc,min} + (n-1)\Delta f, \tag{7}$$

where $f_{sc,n}$ ($n$ = 1, 2, 3, …, $N$) is the frequency of the subcarriers.

By separating pulses from different channels using digital filters centered at the above frequencies, the time-domain low-speed pulses from different channels can be used for obtaining the STFT of the SUT.

## 2.2 Performance improvement introduced by channelization

In our previous work in [21], it is shown that in the case of single channel measurement, the system time resolution is equal to the sweep period $T$ while the system frequency resolution is determined by the sweep chirp rate $k$ and decreases with the increase of $k$. In addition, the relationship between the two parameters and the single channel sweep bandwidth $f_s$ can be expressed as

$$f_s = kT. \tag{8}$$

As shown in Eq. (8), the time resolution and frequency resolution of a single-channel system are mutually constrained, and ultimately limited by the single-channel analysis bandwidth. The smaller the single-channel analysis bandwidth, the better the overall frequency resolution and time resolution. In this work, within each measurement channel, the relationship given in (8) still holds. However, because of the introduction of multiple channels, the analysis bandwidth of the proposed system can be further given as

$$B = \delta_1 = Nf_s = NkT. \tag{9}$$

Eq. (9) indicates that when $f_s$ is fixed, the channelized STFT scheme proposed in this work can greatly expand the instantaneous analysis bandwidth of the system to $N$ times the single channel analysis bandwidth $f_s$. Furthermore, when the instantaneous analysis bandwidth $B$ of the system is fixed, the single channel analysis bandwidth $f_s$ requirements can be reduced by introducing channelized measurement, thereby improving the overall frequency resolution and time resolution of the system when $k$ and $T$ decrease. More generally, by introducing channelized measurement and using a larger number of channels $N$, the instantaneous analysis bandwidth $B$ of the system can be increased while reducing the sweep chirp rate $k$ and sweep period $T$, which means that the proposed STFT scheme can simultaneously improve the analysis bandwidth, time resolution, and frequency resolution compared with our previous STFT work in [21].

## 3. Experimental results and discussion

### 3.1 Experimental setup

An experiment is carried out to verify the proposed channelized analog microwave STFT system, with the experimental setup shown in Fig. 3. The three OFCs in the experiment are generated by modulating CW light waves using single-tone RF signals. In the experiments, the number of channels is set to 3 or 5, so only one Mach–Zehnder modulator (MZM) is used in each OFC module.

A 15.5-dBm optical carrier from LD1 (ID Photonics CoBriteDX1-1-C-H01-FA) is modulated at MZM1 (Fujitsu FTM-7938EZ) by a single-tone RF signal centered at $f_{RF1}$ to generate an OFC. The RF signal is generated from a microwave signal generator (MSG, HP 83752B). Before it is injected into MZM1, the single-tone RF signal is amplified by an electrical amplifier (Multilink MTC5515-751) and then filtered by a bandpass filter. The output of MZM1 is injected into DP-MZM1 (Fujitsu FTM-7961EX), where it is CS-LSSB-modulated by a frequency-sweep signal from port 1 of arbitrary waveform generator 1 (AWG1, Keysight M8195A) to generate the frequency-sweep OFC. The frequency-sweep OFC from DP-MZM1 is amplified by erbium-doped fiber amplifier 1 (EDFA1, Max-Ray EDFA-PA-35-B) and then injected into DP-MZM2 (Fujitsu FTM-7961EX).

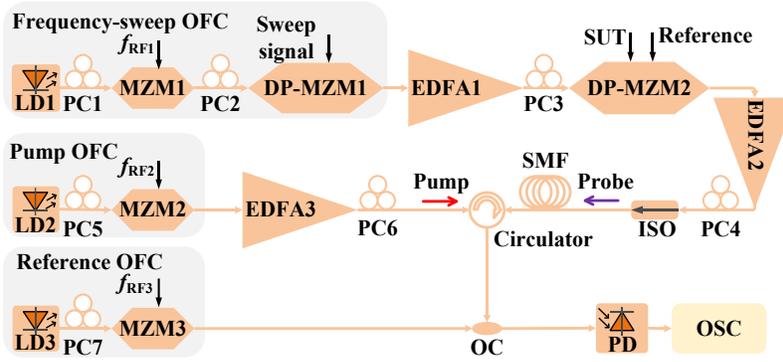

Fig. 3. Experimental setup of the proposed channelized analog microwave STFT system. LD, laser diode; PC, polarization controller; MZM, Mach–Zehnder modulator; DP-MZM, dual-parallel Mach–Zehnder modulator; EDFA, erbium-doped fiber amplifier; SUT, signal under test; OC, optical coupler; ISO, isolator; PD, photodetector; OSC, oscilloscope.

A combined signal, including the SUT and the reference signal, is sent to DP-MZM1 to implement CS-RSSB modulation on the frequency-sweep OFC. Here, the reference signal is generated from port 1 of AWG2 (Keysight M8190A). Due to the limited bandwidth of AWG2, one part of the SUT is generated by up-converting an IF signal from port 2 of AWG2 using an electrical mixer (MITEQ M30). The local oscillator (LO) signal for mixing is generated from port 2 of AWG1 and amplified by an electrical amplifier (CTT ALM/145-5023-293). Because AWG2 can only generate signals within 4 GHz, it can only occupy up to 8 GHz bandwidth after frequency up-conversion. In order to analyze SUTs in a larger bandwidth, in subsequent experiments, the port where AWG2 generates the reference signal simultaneously outputs another part of the SUT in a lower frequency range. The outputs from port 1 of AWG2 and from the mixer are combined in an electrical coupler (Narda 4456-2) and then applied to DP-MZM2 via a 90° hybrid (SHW TH-2/18-3S-90). The output of DP-MZM2 is amplified by EDFA2 (Amonics AEDFA-PTK-DWDM-15-B-FA) and then injected into a 25.2-km single mode fiber (SMF) through an isolator (ISO) as the probe wave. The SMF used here as an SBS medium can be replaced by shorter highly nonlinear optical fibers for ease of practical implementation.

A 15-dBm optical carrier from LD2 (ID Photonics CoBriteDX1-1-C-H01-FA) is modulated at MZM2 (Fujitsu FTM-7938EZ) by a single-tone RF signal centered at $f_{RF2}$ from port 3 of AWG1 to generate the pump OFC. The RF signal is amplified by an electrical amplifier (MITEQ AMF-4B-137157-50-23P-L) before it is applied to MZM2. After being amplified by EDFA3 (Amonics AEDFA-PA-35-B-FA), the pump OFC is reversely sent to the SBS medium through the optical circulator to excite the SBS effect and realize narrowband filtering and amplification of probe waves.

A 13.5-dBm optical carrier from LD3 (ID Photonics CoBriteDX1-1-HC1-FA) is modulated at MZM3 (Fujitsu FTM-7938EZ) by a single-tone RF signal centered at $f_{RF3}$ from MSG2 (MSG,

Agilent 83630B) to generate the reference OFC. The RF signal is amplified by an electrical amplifier (JCA Technology JCA1011-450BC) before being applied to MZM3. The reference OFC and the optical signal after SBS interaction are coupled and then detected in a PD (Nortel PP-10G). The generated electrical signal is monitored by an oscilloscope (OSC, R&S RTO2032).

Multiple polarization controllers (PC1, PC2, PC3, PC5, and PC7) are used to optimize the polarization states of light waves before entering different modulators while PC4 and PC6 are used to obtain the optimal state of the SBS interaction. The Brillouin frequency shift $f_{SBS}$ is around 10.8 GHz and the 3-dB bandwidth of the SBS gain spectrum is around 20 MHz.

## 3.2 Channelized STFT verification with an improved analysis bandwidth

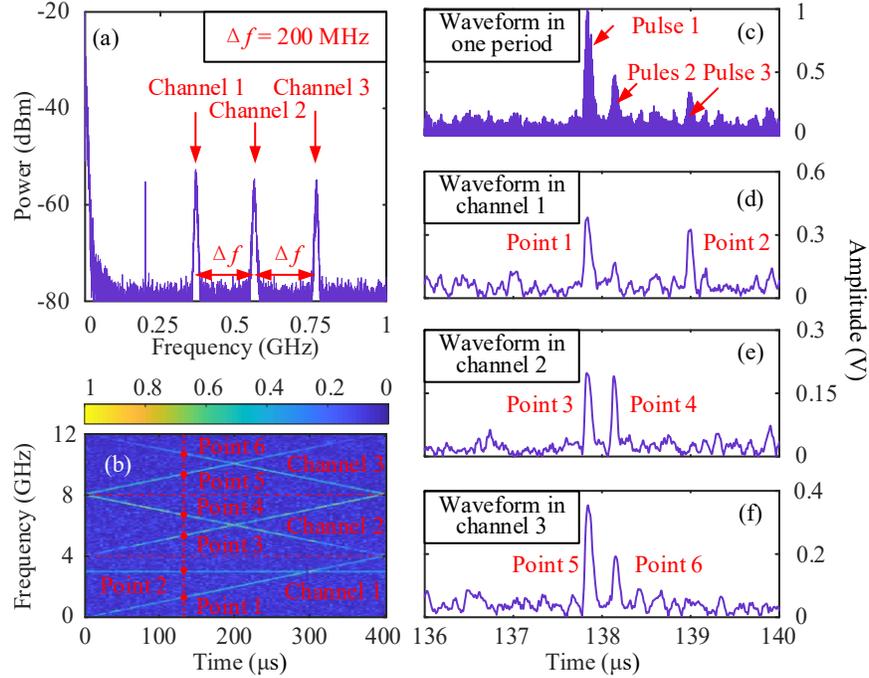

Fig. 4. (a) Electrical spectrum of the signal output from the PD. (b) Measured time-frequency diagram of the SUT (upconverted DLFM signal from 4 GHz–12 GHz and LFM signal from 0–4 GHz) and a 3-GHz reference signal. (c) A period of a waveform from 136 μs–140 μs obtained directly from the PD. Pulse waveforms for STFT in (d) Channel 1, (e) Channel 2, and (f) Channel 3 obtained by filtering and envelope processing the signal in (c) using different subcarriers in (a).

Firstly, the proposed system is demonstrated when $N=3$ and $f_s=4$ GHz, i.e., the system has 3 channels and 12 GHz instantaneous analysis bandwidth. In the experiment, the frequencies of the optical carriers from LD1, LD2, and LD3 are set to 193.2964 THz, 193.3112 THz, and 193.29984 THz. The frequencies of the RF signals ($f_{RF1}$, $f_{RF2}$, and $f_{RF3}$) in the three OFC modules are set to 12 GHz, 16 GHz, and 15.8 GHz. The period $T$, bandwidth $f_s$, and center frequency of the frequency-sweep signal in the frequency-sweep OFC module are set to 4 μs, 4 GHz, and 2 GHz. Under the above settings, the parameters of the system as shown in Fig. 2 conform to Eqs. (2)–(7).

The SUT and reference signal are generated as follows: 1) A linearly frequency-modulated (LFM) signal ranging from 0–4 GHz and having a time length of 400 μs is generated along with a single-tone signal at 3 GHz from port 1 of AWG2; 2) A dual-chirp LFM (DLFM) signal also ranging from 0–4 GHz and having a time length of 400 μs is generated from port 2 of AWG2 as an IF signal, which is then mixed with an 8-GHz LO signal to generate two upconverted DLFM signals from 4 GHz–8 GHz and 8 GHz–12 GHz; 3) The two signals generated above are combined and the LFM signal from 0–4 GHz and the DLFM signals from 4 GHz–12 GHz are used as the SUT while the 3-GHz single-tone signal is used as the reference signal. For the convenience of comparison, the time length of the SUTs in this article is all set to 400 μs, and will not be changed in the subsequent experiments. The reference signal frequency is set to 3 GHz here, which is much higher than that in our previous work in [21], [22], [24] to make the reference frequency more clear in the time-frequency diagram. In practical applications, the reference signal frequency can be set to a much lower frequency as in [21], [22], [24].

As shown in Fig. 4(a), the electrical spectrum of the signal from the PD monitored by an electrical spectrum analyzer (ESA, R&S FSP-40) has three frequency components, which correspond to the pulses carried by three subcarriers in the three channels. The frequencies of the three subcarriers are determined by the settings of the three OFCs and comply with Eqs. (6) and (7). The signal from the PD is captured by the OSC. After filtering and selecting the three frequency components and getting their envelopes respectively, the obtained electrical pulses from the three channels can be obtained. These pulses are further combined to generate the time-frequency information of the SUT and realize the STFT of the SUT, as shown in Fig. 4(b). The frequency resolution is around 60 MHz. It can be seen that the time-frequency information of the SUT and reference signal described above is given correctly, so the time-frequency analysis capability of the channelized analog microwave STFT system is verified. It can be observed that in the time-frequency diagram, the intensity from 4 GHz–5 GHz and 11 GHz–12 GHz is weak. The reason is that the mixer IF bandwidth is 0–3 GHz, and the 4 GHz–5 GHz and 11 GHz–12 GHz signal upconverted by the signal in 3 GHz–4 GHz has a decreased signal intensity.

The waveform in one sweep period from 136 μs–140 μs is further given and shown in Fig. 4(c). Multiple pulses carried by different subcarriers are mixed together. The pulses generated in the three channels after filtering and enveloping are shown in Fig. 4(d)–(f). It can be observed that the six points marked in Fig. 4(b) correspond to six different pulses distributed in the three different channels in Fig. 4(d)–(f). By connecting the pulses of the three channels in channel order and placing them in the position shown by the dashed line in the time-frequency plane, the time-frequency information of the SUT from 136 μs–140 μs is obtained. The complete time-frequency diagram of the SUT can be obtained by performing the same processing on the pulses generated in the three channels for each sweep period. From the above analysis, it can be clearly seen that we can distinguish the measurement results of different channels well through subcarriers while obtaining the time-frequency relationships in different channels and obtaining the time-frequency information of the entire test frequency band

To further validate the time-frequency analysis capability of the proposed channelized STFT system, the non-linearly frequency-modulated (NLFM) signal, LFM signal, frequency-hopping signal, and step-frequency signal all with a frequency range from 0–4 GHz and a 400-μs time length are selected as the IF signal and output from port 2 of AWG2. The IF signal is also upconverted to 4 GHz–12 GHz using an 8-GHz LO signal. An NLFM signal or an LFM signal output from port 1 of AWG2 also with a frequency range from 0–4 GHz is combined with the upconverted signals as SUTs. A 3-GHz single-tone signal is also generated from port 1 of AWG2 as the reference signal.

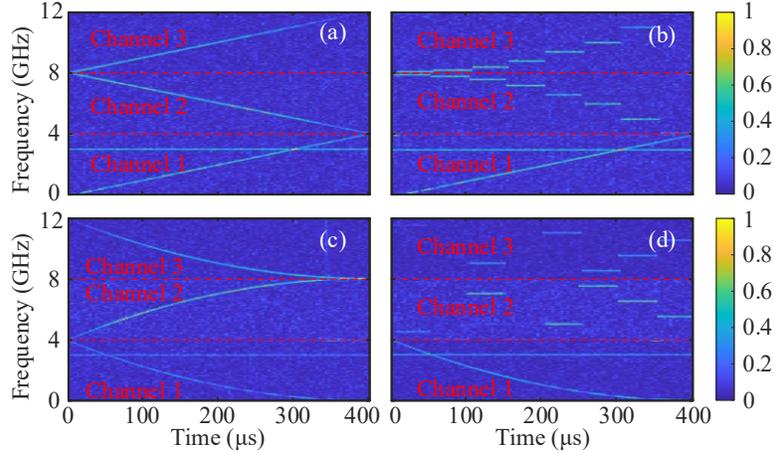

Fig. 5. Measured time-frequency diagrams by the proposed channelized STFT system. The SUTs are (a) LFM signal (0–4 GHz) and upconverted LFM signal (4 GHz–12 GHz), (b) LFM signal (0–4 GHz) and upconverted step-frequency signal (4 GHz–12 GHz), (c) NLFM signal (0–4 GHz) and upconverted NLFM signal (4 GHz–12 GHz), (d) NLFM signal (0–4 GHz) and upconverted frequency-hopping signal (4 GHz–12 GHz). The reference signal is a 3-GHz single-tone signal.

Under these circumstances, the time-frequency diagrams of the four SUTs are shown in Fig. 5. It can be seen that different signal formats can be easily distinguished from the time-frequency diagrams. The time resolution is 4 μs, which equals to the sweep period of the frequency-sweep signal. The frequency resolution is around 60 MHz, which is also consistent with the results given in our previous works in [21] and [24] under a sweep rate of 1 GHz/μs. Nevertheless, in [21] and [24], the instantaneous analysis bandwidth is strictly limited by and equals to the sweep bandwidth of the frequency-sweep signal. In this work, due to the introduction of three OFCs and the first employment of channelized measurement in analog microwave STFT, a measurement bandwidth of 12 GHz is achieved by using a frequency-sweep signal with 4 GHz bandwidth and 3 channels while maintaining the same time and frequency resolution.

As given in Eq. (9), the analysis bandwidth of the proposed system is jointly determined by the sweep bandwidth $f_s$ and the number of channels $N$. By expanding the number of channels via OFCs with more comb lines, the system analysis bandwidth can be further increased while keeping the time and frequency resolution unchanged. The time and frequency resolution of the system can also be adjusted by changing the sweep period $T$ and sweep chirp rate $k$ of the frequency-sweep signal. If $T$ or $k$ is changed and $N$ is fixed, the instantaneous analysis bandwidth will also change with the time and frequency resolution. By adjusting the relative position and spectral line spacing of the OFCs and changing the $f_s$, the system analysis bandwidth $\delta_1$ can also be adjusted. In the following work, the improvement of system analysis bandwidth, time resolution, and frequency resolution by channelization is further demonstrated and analyzed.

### 3.3 System tunability and performance improvement

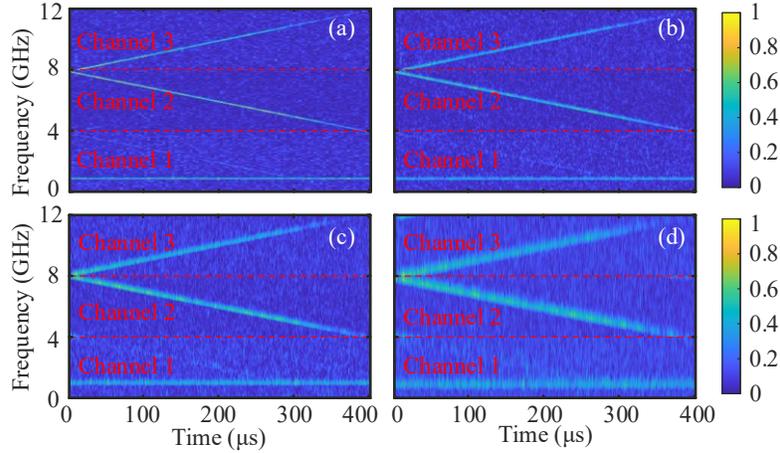

Fig. 6. Measured time-frequency diagrams of the upconverted LFM signal (4 GHz–12 GHz) and the 1-GHz single-tone reference signal when the sweep chirp rate $k$ of the frequency-sweep signal is (a) 1 GHz/µs, (b) 2 GHz/µs, (c) 4 GHz/µs, and (d) 8 GHz/µs.

Next, the system tunability and performance improvement it is further investigated. Firstly, the trade-off between the time resolution and frequency resolution is demonstrated. An IF LFM signal from 0–4 GHz is upconverted to 4 GHz–12 GHz and used as the SUT, and a reference frequency is set to 1 GHz in this study. The system parameters ($f_s$, $N$, $B$) are all kept unchanged as used above except the sweep chirp rate $k$ and the sweep period $T$. According to Eqs. (8) and (9), when the analysis bandwidth is fixed, $k$ and $T$ are inversely proportional. Here, the sweep bandwidth $f_s$ is 4 GHz, so the sweep chirp rate $k$ is set to 1 GHz/µs, 2 GHz/µs, 4 GHz/µs, and 8 GHz/µs when sweep period $T$ is set to 4 µs, 2 µs, 1 µs, and 0.5 µs, respectively. Under these circumstances, the time-frequency diagram of the SUT and the reference signal is shown in Fig. 6. With the increase of $k$ (decrease of $T$), the frequency resolution decreases while the time resolution increases, which is consistent with the conclusion in [21]. However, the advantage of this system is also very clear compared with [21]: although the time resolution and frequency resolution are also mutually constrained, due to the introduction of channelized measurement, the instantaneous analysis bandwidth can be expanded by $N$ times at the same time and frequency resolution.

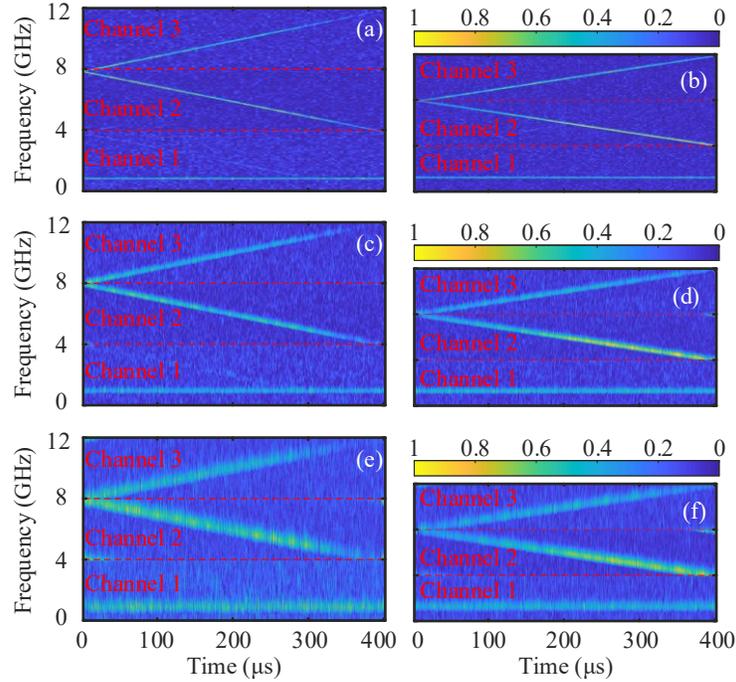

Fig. 7. Measured time-frequency diagrams of the SUT and reference signal when the sweep chirp rate $k$ of the frequency-sweep signal is (a) (b) 1 GHz/μs, (c) (d) 4 GHz/μs, (e) 10 GHz/μs, and (f) 7.5 GHz/μs, and the instantaneous analysis bandwidth is (a) (c) (e) 12 GHz and (b) (d) (f) 9 GHz.

Then, the instantaneous analysis bandwidth of the system is changed from 12 GHz to 9 GHz when $N$ is fixed to 3. Fig. 7 gives a comparison of the analysis results under the two different analysis bandwidths. When the bandwidth is 12 GHz, the sweep chirp rate $k$ is 1 GHz/μs, 4 GHz/μs, and 10 GHz/μs, corresponding to a sweep period $T$ of 4 μs, 1 μs, and 0.4 μs. When the bandwidth is 9 GHz, the sweep chirp rate $k$ is 1 GHz/μs, 4 GHz/μs, and 7.5 GHz/μs, corresponding to sweep period $T$ of 3 μs, 0.75 μs, and 0.4 μs. The SUT is also generated by upconverting an IF LFM signal, which owns a frequency range from 4 GHz–12 GHz or 3 GHz–9 GHz. The reference frequency is set to 1 GHz. By comparing Fig. 7(a) and (b), as well as Fig. 7(c) and (d), it can be seen that while maintaining a constant $N$ (number of channels) and $k$ (sweep chirp rate and frequency resolution), reducing the maximum analysis bandwidth can result in superior time resolution. Additionally, Fig. 7(e) and (f) reveal that decreasing the maximum analysis bandwidth while holding $T$ (sweep period and time resolution) and $N$ (number of channels) constant can enhance the frequency resolution of the system. It is worth noting that, because the sweep range of the IF LFM signal is 3 GHz and not limited by the working frequency of the mixer, there will be no power attenuation at both ends of the dual-chirp LFM signalin Fig. 7(b), (d), and (f). From the above results, the conclusion discussed below Eq. (9) is verified.

It is also observed from Fig. 7 that, when a frequency component appears near the starting or stopping frequencies of a channel, that is, when pulses are generated at the boundary of a sweep period, an undesired frequency will be shown on the opposite side of the channel besides the correct analysis results. This phenomenon can be explained as follows: By repeating periodic frequency sweeping, the frequency of the SUT is mapped to time-domain pulses, and the frequency to be measured is determined by the pulse position; At the time of switching between two adjacent sweep periods, the frequency being measured will switch between the maximum and minimum values; If the frequency to be measured appears around the switching area, the pulse generated by it will spread to the adjacent period due to its certain width, resulting in frequencies that should not appear in the adjacent period as shown in Fig. 7. Essentially, this phenomenon is caused by the pulse width of the generated pulses. The wider the generated pulses (the worse the frequency resolution), the more obvious the above phenomenon is.

By adding a certain amount of idle time before and after each sweep period, the above phenomenon can be suppressed. The cost is that the effective sweep time is reduced, and the overall performance of the system will be reduced to a certain extent. When the effective sweep time is reduced, two cases exist: 1) To achieve the same analysis bandwidth, the sweep chirp rate $k$ needs to be increased, and the system frequency resolution will decrease; 2) To achieve the same frequency resolution, it is necessary to ensure that the sweep chirp rate $k$ remains unchanged, and in this case, the analysis bandwidth will decrease.

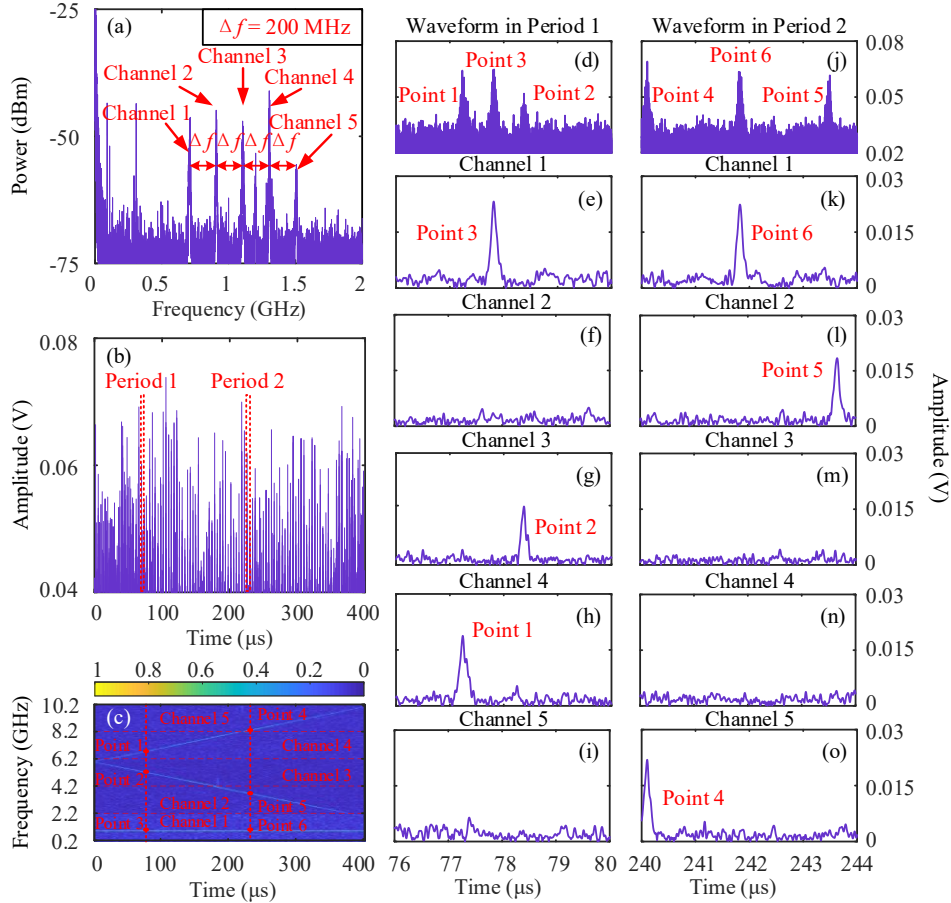

Fig. 8 (a) Electrical spectrum of the signal output from the PD. (b) Waveform of the signal from the PD. (c) Measured time-frequency diagrams of the SUT (upconverted LFM signal from 2 GHz–10 GHz) and a 1-GHz reference signal. Waveforms after PD and in each channel in two sweep periods (d)–(i) from 76 μs–80 μs and (j)–(o) from 240 μs–244 μs.

Then, the proposed channelized analog microwave STFT system is further investigated when the number of channels $N$ is increased to 5. In this case, the spectral line number is also 5. The amplitudes of the RF signals applied to the three MZMs and the bias points of the three MZMs are jointly adjusted to generate the three OFCs with 5 comb lines. The sweep bandwidth $f_s$ is changed to 2 GHz, so the instantaneous analysis bandwidth $B$ of the system is 10 GHz. Here, instantaneous analysis bandwidth is only set to 10 GHz because there is no suitable mixer in our laboratory to produce a proper SUT with higher frequencies. The system parameter configuration is as follows: The frequencies of the optical carriers from LD1, LD2, and LD3 are changed to 193.2962 THz, 193.3112 THz, and 193.2993 THz; The frequencies of the RF signals ($f_{RF1}$, $f_{RF2}$, and $f_{RF3}$) in the three OFC modules are changed to 10 GHz, 12 GHz, and 11.8 GHz; The sweep period $T$ of the frequency-sweep signal is 4 μs and the sweep chirp rate $k$ is 0.5 GHz/μs. In this case, the analysis frequency range is from 0.2 GHz–10.2 GHz.

An LFM signal with a frequency range from 0–4 GHz is used as the IF signal and the frequency of the LO signal is 6 GHz, so the signal after frequency upconversion is a DLFM signal with a frequency range from 2 GHz–10 GHz. The frequency of the single-tone reference signal is set to 1 GHz. The measurement results are given in Fig. 8. As shown in Fig. 8(a), the electrical spectrum of the generated pulse signal from the PD has five frequency components, each corresponding to one of the five channels. The frequencies of the five subcarriers are determined by the settings of the three OFCs and comply with Eqs. (6) and (7). The corresponding waveform captured by the OSC is shown in Fig. 8(b). The time-frequency diagram of the SUT and the reference signal obtained from the waveform is shown in Fig. 8(c). As can be seen, the DLFM signal and the reference signal are given correctly. Two sweep periods of the waveform from 76 μs–80 μs and 240 μs–244 μs marked in Fig. 8(c) are given in Fig. 8(d) and (j). Multiple pulses carried by different subcarriers are mixed together. The signals generated in the five channels after filtering and enveloping are shown in Fig. 8(e)–(i) and Fig. 8(k)–(o). It can be observed that the three points marked in each sweep period in Fig. 8(c) correspond to three different pulses distributed in the three of the five channels in Fig. 8(e)–(i) and Fig. 8(k)–(o).

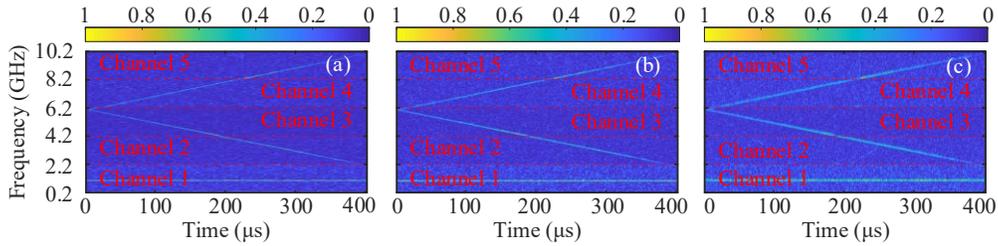

Fig. 9. Measured time-frequency diagrams of the SUT and the reference signal when the sweep chirp rate $k$ of the frequency-sweep signal is (a) 0.5 GHz/μs, (b) 1 GHz/μs, and (c) 2 GHz/μs.

Then, the system is configured with different sweep chirp rate $k$ (0.5 GHz/μs, 1 GHz/μs, and 2 GHz/μs) and sweep period $T$ (4 μs, 2 μs, and 1 μs) while maintaining the sweep bandwidth $f_s$ at 2 GHz. The SUT and reference signal are the same as that used in obtaining Fig. 8 and the measured time-frequency diagrams are shown in Fig. 9. The results in Fig. 9(a), (b), and (c) can be compared with that in Fig. 6(a), (b), and (c). By selecting one channel from the time-frequency diagrams in Fig. 6(a), (b), and (c), and two adjacent channels from that in Fig. 9(a), (b), and (c), we can compare the two cases with the same analysis bandwidth (4 GHz) but different number of channels ($N = 1$ in Fig. 6 and 2 in Fig. 9) to show enhancement of the system resolution through increasing the number of channels. By comparing Fig. 6(a) and Fig. 9(a), Fig. 6(b) and Fig. 9(b), or Fig. 6(c) and Fig. 9(c), it can be clearly seen that when the sweep period $T$ and the analysis bandwidth remain constant, increasing the number of channels $N$ can improve the frequency resolution of the system. By comparing Fig. 6(a) and Fig. 9(b), it is also found that when the sweep chirp rate $k$ and the analysis bandwidth remain constant, increasing the number of channels $N$ can also improve the time resolution of the system. Therefore, by increasing the number of channels $N$, the time resolution and frequency resolution of the system can be improved, and the conclusion given below Eq. (9) is further verified.

**3.4 Discussion and comparison**

In our previous work in [21], the frequency resolution (determined by sweep chirp rate $k$) and time resolution (determined by the sweep period $T$) of the STFT are mutually constrained and limited by the analysis bandwidth of the system (also the sweep bandwidth of the frequency-sweep signal $f_s$) by $f_s = kT$. Due to local stationarity being the fundamental premise of STFT, to ensure good signal stationarity of the SUT within a sweep period, the sweep chirp rate $k$ must be fast enough. For example, in [21] and this work, $k$ is as high as several GHz/μs to ensure a μs level time resolution and an analysis bandwidth of several GHz. Nevertheless, such a high-speed frequency-sweep signal is very difficult to be generated in the electrical domain. The high-speed electrical frequency-sweep signal in [21] is generated by an AWG, which is very expensive and is difficult to integrate directly as a module into miniaturized measurement systems. A method implemented by directly modulating a distributed feedback laser using a sawtooth waveform is proposed in [23], which avoids the use of the high-speed electrical frequency-sweep source. However, the optical method in [23] is also limited in the generation of frequency-sweep optical signal with even higher sweep bandwidth.

In this work, for the first time, channelized measurement is introduced to the analog microwave STFT system, and the number of channels $N$ provides a new dimension for system optimization. It can greatly reduce the sweep bandwidth $f_s$ of the required frequency-sweep signal to the instantaneous analysis bandwidth $B$ divided by the number of channels $N$. In other words, if the sweep bandwidth $f_s$ is the same as that in [21], the instantaneous analysis bandwidth $B$ can be increased by $N$ times. In addition, channelization can also be used to improve the time and frequency resolution of the STFT system. As given in Eq. (9), since $B = \delta_1 = Nf_s = NkT$, the instantaneous analysis bandwidth $B$, the sweep chirp rate $k$, and the sweep period $T$ can be simultaneously improved by increasing the number of channels $N$. If $N$ is represented as

$$N = N_1 \times N_2 \times N_3. \tag{10}$$

When the STFT system has $N$ channels, if the sweep chirp rate $k$ is changed to $k/N_2$, the sweep period $T$ is changed to $T/N_3$, and the instantaneous analysis bandwidth can be written as

$$B' = N \times \frac{k}{N_2} \times \frac{T}{N_3} = \frac{N}{N_2 N_3} kT = N_1 kT \tag{11}$$

As can be seen from Eq. (11), $N_1$ is used to increase the instantaneous analysis bandwidth by $N_1$ times, $N_2$ is used to decrease the sweep chirp rate by $N_2$ times (increase the frequency resolution), and $N_3$ is used to decrease the sweep period by $N_3$ times (increase the time resolution by $N_3$ times). In the experiment in this article, STFT with 3 channels or 5 channels is demonstrated, and the above analysis is well confirmed by the experimental results. By further increasing the number of channels $N$, the proposed STFT system can further reduce the bandwidth requirement for the frequency-sweep electrical signals to several hundred MHz or even lower, while achieving much greater instantaneous analysis bandwidth and better frequency and time resolution at the same time.

Another issue that needs to be addressed here is the coherence between the three OFCs. As shown in the experimental setup in Fig. 3, the three OFCs are generated from three different OFC modules, each consisting of an independent laser and an MZM. In the experiment, the three OFCs are not locked together. Due to the need to accurately set the OFC frequencies in the system, the long-term stability in the experiment is limited. Indeed, the carriers for the three OFCs can be generated by using a single LD with different frequency shift operations or optical injection locking, so the stability of the system will no longer be a problem. In the article, a proof-of-concept experiment is given to verify the proposed method, so three independent LDs are used to simplify the experiment. It is also found that in short-term operations, the stability and obtained time-frequency diagrams are as good as that in our previous work.

Table 1. Comparison of Different Photonics-Assisted STFT Methods

| | Techniques /Devices Used | Dispersion Dependent | Analysis Bandwidth | Frequency Resolution | Time Resolution | Reconfigurability | Number of Channels |
|---|---|---|---|---|---|---|---|
| Ref. [18] | LCFBG Array | Yes | Unspecified | Unspecified | Unspecified | Low | / |
| Ref. [19] | Time Sampling & DCF | Yes | 2.43 GHz | 340 MHz | 5 ns | Middle | / |
| Ref. [20] | Time Lens & DCF | Yes | 1.98 GHz | 60 MHz | 6.25 ns | Middle | / |
| Ref. [21] | SBS-FTTM | No | $kT$ (12 GHz $^a$) | Determined by $k$ (60 MHz@4 GHz analysis bandwidth) | $T$ (0.5 μs) | High | 1 |
| This work | SBS-FTTM | No | $N_1kT$ (12 GHz $^a$) | Determined by $k/N_2$ (35 MHz@10 GHz analysis bandwidth $^b$) | $T/N_3$ (0.4 μs $^c$) | Very High | $N=N_1×N_2×N_3$ (3 or 5) |

$^a$ The analysis bandwidth of 12 GHz is achieved in this work and Ref. [21]. However, the two works use frequency-sweep signals with different bandwidths: 4 GHz in this work and 12 GHz in Ref. [21].
$^{b,c}$ Demonstrated in Fig. 9(a)$^b$ and Fig. 7(e)$^c$ in this work, which can be further increased by increasing the number of channels.

Finally, a comparison between this work and the previously reported photonics-assisted STFT systems is given in Table 1. The method proposed in this work has experimentally verified the maximum instantaneous analysis bandwidth, and the bandwidth can also be significantly expanded by increasing the number of channels. The time resolution and frequency resolution can also far exceed the STFT system we previously reported in [21], as discussed above, by expanding the number of channels. Compared with other dispersion-based methods, the analysis bandwidth and frequency resolution of the system has huge advantages. Although the time resolution is not comparable to the results in [19, 20], by employing more channels, the time resolution can be much improved.

## 4. Conclusion

In summary, we have proposed a photonics-assisted analog microwave STFT system with improved measurement performance by employing channelization measurement. To the best of our knowledge, it is the first time that channelization measurement technology is introduced in the STFT system based on frequency sweeping and FTTM. The channelization provides a new dimension for system performance optimization, which can simultaneously improve the instantaneous analysis bandwidth, frequency resolution, and time resolution of the STFT system. A proof-of-concept experiment is carried out. 12-GHz and 10-GHz analysis bandwidth is implemented by using a 4-GHz frequency-sweep signal and 3 channels and a 2-GHz frequency-sweep signal and 5 channels. Measurement performance improvement is well confirmed. By using more channels, the method proposed in this article will greatly advance the practical application of the STFT system based on frequency sweeping and FTTM.

**Acknowledgements**
This work was supported in part by National Natural Science Foundation of China (61971193), in part by Natural Science Foundation of Shanghai (20ZR1416100), in part by SongShan Laboratory Pre-research Project (YYJC072022006), in part by Shanghai Aerospace Science and Technology Innovation Fund (SAST2022-074), and in part by Science and Technology Commission of Shanghai Municipality (22DZ2229004).